\documentclass[prl,twocolumn,showpacs,groupedaddress]{revtex4}
\usepackage{graphicx}
\usepackage{epsfig,color}
\usepackage{dcolumn}

\begin{document}

\title{Feedback spin resonance in superconducting CeCu$_2$Si$_2$ and CeCoIn$_5$}

\author {I.~Eremin$^{1,2}$}
\author {G. Zwicknagl$^{2}$}
\author {P.~Thalmeier$^{3}$}
\author {P.~Fulde$^{1}$}
\affiliation {$^1$Max-Planck-Institut f\"{u}r Physik komplexer
Systeme, D-01187 Dresden, Germany} \affiliation {$^2$Institute
f\"{u}r Mathematische und Theoretische Physik, TU Braunschweig,
D-38106 Braunschweig, Germany} \affiliation {$^3$Max-Planck Institut
f\"ur Chemische Physik fester Stoffe, D-01187 Dresden, Germany}

\date{15.04.2008}

\begin{abstract}
We show that the recently observed spin resonance modes in
heavy-fermion superconductors CeCoIn$_5$ and CeCu$_2$Si$_2$ are
magnetic excitons originating from superconducting quasiparticles.
The wave vector ${\bf Q}$ of the resonance state leads to a powerful
criterion for the symmetry and node positions of the unconventional
gap function. The detailed analysis of the superconducting feedback
on magnetic excitations reveals that the symmetry of the
superconducting gap corresponds to a singlet $d_{x^2-y^2}$ state
symmetry in both compounds. In particular this resolves the
long-standing ambiguity of the gap symmetry in CeCoIn$_5$. We
demonstrate that in both superconductors the resonance peak shows a
significant dispersion away from ${\bf Q}$ that should be observed
experimentally. Our results suggest a unifying nature of the
resonance peaks in the two heavy-fermion superconductors and in
layered cuprates.
\end{abstract}


\maketitle

The relation between unconventional superconductivity and magnetism
in heavy-fermion systems and doped transition metal oxides is one of
the most interesting research areas in condensed matter physics.
Despite of certain differences  concerning the proximity to a Mott
insulator in the transition metal oxides and the weak hybridization
of the $f$-electrons in the heavy-fermion systems, it is widely
believed that in both cases short-range antiferromagnetic (AF) spin
fluctuation are responsible for Cooper-pairing with a $d$-wave order
parameter. Furthermore, unconventional superconductivity itself has
a strong feedback on the magnetic spin excitations in these systems
below the superconducting transition temperature $T_c$. One example
is the famous resonance peak observed in high-T$_c$ cuprates by
means of inelastic neutron scattering(INS) \cite{rossat} whose
nature is still actively debated\cite{palee}. The complexities of
the cuprates including the electronic inhomogeneities or stripe
formation complicates here the interpretation of the experimental
data.

For some time the cuprates were considered to be unique in
displaying the resonance peak in the superconducting state. Then,
remarkably INS revealed the formation of a new magnetic mode in the
superconducting state of the 5$f$-heavy fermion compound
UPd$_2$Al$_3$ with T$_c$ = 1.8 K \cite{sato}. Its sharply peaked
intensity, its temperature dependence and energy position well below
2$\Delta_0$ (with $\Delta_0$ being the maximum of the
superconducting gap) strongly resembles the resonance peak seen in
high-T$_c$ cuprates. In distinction to the cuprates, the symmetry of
the gap function was unclear for UPd$_2$Al$_3$. Thermal conductivity
measurements in a rotating magnetic field showed that it has nodes
in the hexagonal plane but could not determine the symmetry uniquely
\cite{watanabe04}. As already noticed in \cite{sato,bernhoeft00} and
shown explicitly in \cite{chang07} the observation of a resonance at
the AF wavevector {\bf Q} puts a stringent condition on the gap
symmetry by requiring a sign change $\Delta_{\bf k + Q}=-\Delta_{\bf
k}$ under translation by {\bf Q}. Together with thermal conductivity
results this determines unambiguously the gap function as
$\Delta_{\bf k}=\Delta_0\cos ck_z$ which has node lines at the
boundaries of the antiferromagnetic Brillouin zone. The case of
UPd$_2$Al$_3$ shows not only that the magnetic resonance due to a
feedback effect of the superconducting state is a universal
phenomenon in unconventional superconductors but also, that
inelastic neutron scattering (INS) is a powerful technique to
determine the superconducting gap symmetry.

Indeed, very recently the resonance peak has been also observed
below the superconducting transition temperatures in Ce-based
heavy-fermion compounds, namely in CeCu$_2$Si$_2$\cite{stockert} at
the (incommensurate) wave vector {\bf Q}$_{SDW} \approx (0.22 \cdot
2\pi/a, 0.22 \cdot 2\pi/a, 0.53 \cdot 2\pi/c)$ and in
CeCoIn$_5$\cite{broholm} at the antiferromagnetic wave vector {\bf
Q}$_{AF} \approx (\pi/a,\pi/a,\pi/c)$ . The latter system which has
quasi-two-dimensional tetragonal crystal structure shows the highest
superconducting transition temperature $T_c =2.7$K among
heavy-fermion compounds\cite{petrovic}. Its gap symmetry has been
long disputed because of conflicting results from angle resolved
magnetothermal conductivity \cite{izawa01} ($d_{x^2-y^2}$ -symmetry)
and specific heat \cite{aoki} ($d_{xy}$-symmetry) measurements. The
INS  results \cite{broholm} and the analysis presented here give a
clear resolution of this puzzle in favor of the $d_{x^2-y^2}$  state
which underlines again the importance of INS for determining the
superconducting gap symmetry. Heavy quasiparticles in the Ce
compounds \cite{koitzsch,fujimori} of predominantly 4$f$ character
are more strongly correlated than 5$f$ quasiparticles or
quasiparticles in cuprates. Therefore, similarities of the resonance
state indicate a generic feature.

Here, we analyze the dynamical magnetic susceptibility in
CeCu$_2$Si$_2$ and CeCoIn$_5$ below the superconducting transition
temperature, assuming unconventional character of the Cooper-pairing
originating from the exchange of the antiferromagnetic spin
fluctuations. We show that in both cases the resonance feature
evolves at the magnetic instability wave vector. We discuss the
dispersion of the resonant excitations as a function of the momentum
and demonstrate that in both compounds the $d_{x^2-y^2}$-wave
symmetry of the superconducting order parameter is consistent with
the INS experiments.

\begin{figure}[h]
\includegraphics[angle=0,width=0.9\linewidth]{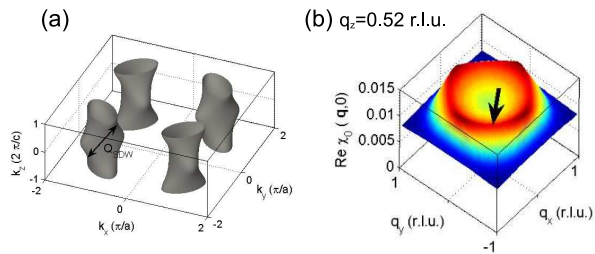}
\caption{(a) Calculated Fermi surface for the main electron sheet in
CeCu$_2$Si$_2$ using the tight-binding parametrization in the form $
\varepsilon_{\bf k} = 2t_1 \left(\cos k_x a +\cos k_y a \right) + 4
t_2 \cos k_x a \cos k_y a +8t_3 \cos \frac{k_xa}{2} \cos
\frac{k_ya}{2}\cos \frac{k_zc}{2} -\mu$ where $t_1 = 17.5$, $t_2 =
-5.2$, $t_3 = -11.2$ and $\mu = -57.4$ (in K) are the hopping
integrals and the chemical potential, respectively. Here, $a=4.1
\AA$ and $c=9.92 \AA$ are the lattice constants. The arrow indicates
the SDW scattering wave vector. (b) Calculated static spin
susceptibility on two-dimensional mesh for $q_z = 0.52$ r.l.u. The
arrow indicates the SDW ordering wave vector, {\bf Q}$_{SDW}$, as
observed in the experiment \protect\cite{zwick92}.} \label{fig1}
\end{figure}

Starting our analysis by considering the CeCu$_2$Si$_2$ system, we
note that a good starting point for calculating the magnetic
susceptibility is the band structure obtained within the
renormalized band theory\cite{zwick01,zwick92,thalmAL}. In this
approach the phase shifts for the $4f$-states are introduced
empirically into the {\it ab-initio} local density approximation
(LDA) scheme to generate heavy quasiparticle bands at the Fermi
level. Crystalline Electric Field (CEF) splitting strongly affects
the quasiparticle dispersion and has to be properly accounted for
\cite{goremichkin}. Note that the formation of a resonance in the
spin response is determined mainly by the unconventional symmetry of
the superconducting order parameter. Therefore it is reasonable to
use a tight-binding parametrization fit to the main heavy
quasiparticle band that crosses the Fermi level. The corresponding
Fermi surface consisting of the stacked columns along $c$ direction
is shown in Fig.\ref{fig1}(a) in the first few Brillouin zones. The
obtained Fermi surface shows a flat part connected by the nesting
wave vector ${\bf Q}_{SDW} = (0.22 \cdot 2 \pi / a, 0.22 \cdot 2 \pi
/a, 0.52 \cdot 2 \pi /c  )$ as indicated by the arrow. As discussed
in earlier work \cite{stock04} the {\bf Q}$_{SDW}$ wavevector agrees
very well with the experimentally observed SDW in the so-called
A-phase of that compound.

In Fig.\ref{fig1}(b) we show the real part of the Lindhard
susceptibility $\chi_0({\bf q})$ calculated with the parameterized
heavy quasiparticles.
%
%
One finds that Re $\chi({\bf q},0)$ is peaked at {\bf Q}$_{SDW}$.
When compared with a fully renormalized band structure
calculation\cite{zwick92} we observe that the tight-binding bands
result in a somewhat more pronounced nesting and yield a whole
contour of the nesting wave vectors around {\bf Q}$_{SDW}$. This is
due to the relatively simple band structure that contains nearest
neighbor hopping integrals only. However, this does not influence
our results concerning the resonance feature below T$_c$, since the
latter is determined mainly by the special nature of the
superconducting gap.

The resonance peak in the SC state of CeCu$_2$Si$_2$ as well as of
CeCoIn$_5$ can be understood by considering the dynamical spin
susceptibility within the random phase approximation (RPA), {\it
i.e.},
\begin{equation}
\chi_{RPA} ({\bf q}, \omega) =  \frac{\chi_0 ({\bf q},
\omega)}{1-U_{\bf q} \chi_0 ({\bf q}, \omega)},
\end{equation}
where $U_{\bf q}$ is the fermionic four-point vertex and $\chi_0
({\bf q}, \omega)$ is the heavy quasiparticle susceptibility. The
latter is given by the sum of the well-known bubble diagram
consisting of either normal or anomalous (T$<$T$_c$) Green
functions. For large momenta {\bf q}, Im$\chi_0 ({\bf q}, 0)$ is
zero at low frequencies and can exhibit a discontinuous jump at the
onset frequency of the p-h continuum $\Omega_c = \min \left(
|\Delta_{\bf k}| + |\Delta_{\bf k+q}| \right)$ where both {\bf k}
and {\bf k+q} lie on the Fermi surface\cite{pines}. Note, however,
that the discontinuity in Im$\chi_0$ occurs only if
$\mbox{sgn}(\Delta_{\bf k}) = - \mbox{sgn}(\Delta_{\bf k+q})$ which
is only possible for unconventional order parameters. A
discontinuity in Im$\chi_0$ leads to a logarithmic singularity in
Re$\chi_0$. As a result, the resonance conditions (i) $U_{\bf
q}\mbox{Re}\chi_0({\bf q}, \omega_{res})= 1$ and (ii) Im$\chi_0({\bf
q}, \omega_{res}) = 0$ can be both fulfilled at $\omega_{res} <
\Omega_c$ for any $U_{\bf q} >0$, leading to the occurrence of a
resonance peak in form of a spin exciton below T$_c$. For finite
quasiparticle damping $\Gamma$, condition (i) can only be satisfied
if $U_{\bf q}
>0$ exceeds a critical value, while condition (ii) is replaced by
Im$\chi_0({\bf q}, \omega_{res}) << 1$.

\begin{figure}[h]
\includegraphics[angle=0,width=0.9\linewidth]{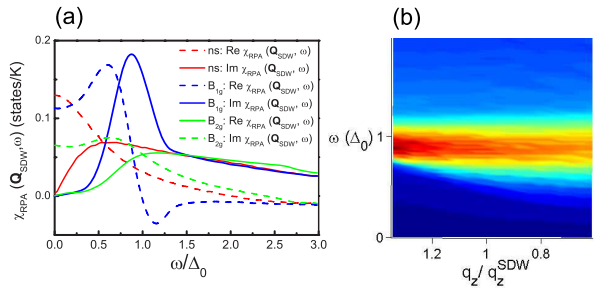}
\caption{(a) Calculated real and imaginary parts of the RPA spin
susceptibility for the normal (red) and superconducting state for
the $B_{1g}$ (blue) ($\Delta_{\bf k} = \Delta_0 \left( \cos k_x a -
\cos k_y a \right)$) and $B_{2g}$ (green) ($\Delta_{\bf k} =
\Delta_0 \sin \frac{k_x a}{2}\sin \frac{k_y a}{2}$)  symmetry of the
superconducting gap. Here, we assume $U_{{\bf Q}_{SDW}} \approx 4
t_1$ to satisfy the resonance condition in the superconducting
state. For the numerical purpose we also set the damping $\Gamma =
2$K. (b) Calculated dispersion of the resonance peak in
CeCu$_2$Si$_2$ for (0.22, 0.22, $q_z$) direction. The dispersion is
nearly flat around {\bf Q}$_{SDW}$ due to the two-dimensional
structure of the superconducting gap. Here, we have introduced an
interaction peaked at {\bf Q}$_{SDW}$ in the form $U_{\bf q} \approx
U_{{\bf Q}_{SDW}}\left[1-B \cdot \frac{\left({\bf
q-Q}_{SDW}\right)^2}{{\bf Q}_{SDW}^2}\right]$ with B=0.5.}
\label{fig2}
\end{figure}

Since the symmetry of the superconducting gap has not yet been
determined unambiguously in CeCu$_2$Si$_2$, we have analyzed all the
spin singlet $s$- and $d$-wave functions allowed by the
crystal-group symmetry of the lattice \cite{zwick92,ozaki,thalmNEW}.
We have found that the resonance-like feature, {\it i.e.}, the
discontinuous jump in Im$\chi_0$ and the corresponding logarithmic
singularity in Re$\chi_0$ occur at the SDW wave vector {\bf
Q}$_{SDW}$ for three types of the order parameters: $\Delta_{\bf k}
= \Delta_0 \left( \cos k_x a - \cos k_y a \right)$ belonging to the
$B_{1g}$ irreducible representation, $\Delta_{\bf k} = \Delta_0 \sin
\frac{k_x a}{2}\sin \frac{k_y a}{2}\cos \frac{k_z c}{2}$ belonging
to the $B_{2g}$ irreducible representation, and also for each of the
two components of the $E_{1g}$ representation, $\Delta_{\bf k} =
\Delta_0 \sin k_x a \sin k_z c$ and $\Delta_{\bf k} = \Delta_0 \sin
\frac{\left( k_x +k_y \right)a}{2} \sin \frac{k_z c}{2}$. An
important finding is that for a given dispersion the resonance in
the $B_{1g}$ channel is by far the strongest. In Fig.\ref{fig2}(a)
we show the RPA susceptibility in the normal and the superconducting
state for the $B_{1g}$ ($\Delta_{\bf k} = \Delta_0 \left( \cos k_x a
- \cos k_y a \right)$) and for the $B_{2g}$ ($\Delta_{\bf k} =
\Delta_0 \sin \frac{k_x a}{2}\sin \frac{k_y a}{2}$) channels,
respectively. One finds that the susceptibility for the $B_{1g}$
symmetry is larger in the superconducting state than it is in the
normal state, while in the $B_{2g}$ channel there is no enhancement
of the normal state spin susceptibility. This definitely points
towards a $d_{x^2-y^2}$-wave symmetry of the superconducting gap in
CeCu$_2$Si$_2$ since a very sharp resonance was found at {\bf
Q}$_{SDW}$ in INS \cite{stockert}.
We note, however, that the magnitude of the resonance for each
symmetry is a result of a competition of various features, {\it
i.e.}, the curvature of the Fermi surface at the points connected by
{\bf Q}$_{SDW}$, the absolute value of the superconducting gap and
also the velocity at the node of the gap. Therefore a modification
of the electronic dispersion, and the actual angular dependence of
the superconducting gap can modify the results for the absolute
intensity of the resonance peak. This is, for example, the case for
the non-monotonic $d-$wave gap in the electron-doped
cuprates\cite{ismer}.

The resonance condition is also satisfied for momenta {\bf Q}$_i$
slightly away from {\bf Q}$_{SDW}$ as long as
$\mbox{sgn}(\Delta_{\bf k}) = - \mbox{sgn}(\Delta_{{\bf k+Q}_i})$.
We find that the resonance is readily suppressed by variation of the
in-plane $(q_x,q_y)$-momentum and exists only in the close vicinity
to ${\bf Q}_{SDW}$. This is because for an incommensurate momentum
there are always scattering processes which involve parts of the
Fermi surface with $\mbox{sgn}(\Delta_{\bf k}) = +
\mbox{sgn}(\Delta_{{\bf k+Q}_i})$ thus suppressing the resonance. In
addition, varying {\bf Q}$_i$ away from {\bf Q}$_{SDW}$ along the
in-plane ($q_x,q_y,q_z^{SDW}$) direction the nesting condition is
also lost which overall yields a decrease of Re$\chi_0$ (see also
Fig.\ref{fig1}(b)). Since the suppression of the resonance occurs
for all symmetries of the above mentioned superconducting order
parameters the resonance peak is confined to momentum
($q_x,q_y,q_z^{SDW}$) in the plane around {\bf Q}$_{SDW}$. At the
same time the situation is less dramatic with respect to the $q_z$
momentum dependence. In 
Fig.\ref{fig2}(b) we show the dispersion of the resonance along the
$q_z$ direction. One finds that it remains nearly flat as one
departs from {\bf Q}$_{SDW}$. As a matter of fact,  for a constant
interaction $U_{\bf q}$, the resonance shows a weak dispersion
resulting from the slight change of the Re$\chi_0$ away from {\bf
Q}$_{SDW}$.
\begin{figure}[t]
\includegraphics[angle=0,width=0.9\linewidth]{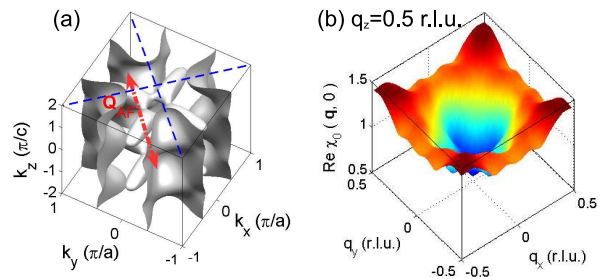}
\caption{(a) Calculated Fermi surface for CeCoIn$_5$ using the band
structure parameters adopted previously\cite{tanaka}. The
dash-dotted arrow points at states at the Fermi surface scattered by
the antiferromagnetic wave vector, {\bf Q}$_{AF}$. The dashed lines
depict the position of nodes in the first BZ for a superconducting
order parameter of $d_{x^2-y^2}$-wave symmetry . Following Ref.
\onlinecite{tanaka} we set the energy unit 0.26eV. (b) Calculated
static spin susceptibility of non-interacting electrons in
CeCoIn$_5$ for $q_z = 0.5$ r.l.u. on a two-dimensional mesh.}
\label{fig3}
\end{figure}

Let us now turn to the resonance peak formation in CeCoIn$_5$
\cite{broholm}. According to the band structure
calculations\cite{tanaka}, CeCoIn$_5$ comprises several $f$- and
conduction bands which are hybridized in a complex manner.
Therefore, it is quite difficult to reproduce a resulting structure
by using a single band model. Introducing a two-band model is much
more appropriate. In fact, this has been done previously for
CeCoIn$_5$ \cite{tanaka} and it has been found that the resulting
energy dispersion crossing the Fermi level can be written as:
\begin{eqnarray}
E_{\bf 2k} = \frac{1}{2}\left[(\varepsilon_{\bf k}^c+E_{\bf
k}^{f})-\sqrt{(E_{\bf k}^{f}-\varepsilon_{\bf k}^c)^2+4V_{\bf
k}^2}\right]
\end{eqnarray}
where $E_{\bf k}^{f}$ and $\varepsilon_{\bf k}^{c} $ is the
effective $f-$band and the conduction band dispersions,
respectively, and $V_{\bf k}$ is the effective hybridization
strength, renormalized by the on-site $f-f$-Coulomb repulsion. The
resulting Fermi surface is shown in Fig. \ref{fig3}(a). Like for
CeCu$_2$Si$_2$ the present Fermi surface has again nesting
properties. However, here it occurs for the the commensurate
antiferromagnetic wave vector {\bf Q}$_{AF} = (\pi/a, \pi/a,
\pi/c)$. This agrees with recent INS data on the normal
state\cite{broholm}. In Fig.\ref{fig3}(b) we show the real part of
the Lindhard spin susceptibility for the normal state calculated on
a two dimensional mesh. In accordance with the Fermi surface
topology we find that the spin susceptibility is peaked at the
antiferromagnetic wave vector, {\bf Q}$_{AF}$. It is important that
these antiferromagnetic spin fluctuations are responsible for the
formation of $d$-wave superconductivity in this compound.
%
%

\begin{figure}[h]
\includegraphics[angle=0,width=0.9\linewidth]{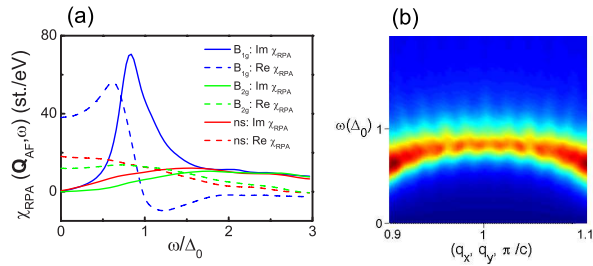}
\caption{(a) Calculated real and imaginary part of the RPA
susceptibility at the antiferromagnetic wave vector {\bf Q}$_{AF}$
for CeCoIn$_5$ as a function of frequency in the normal (red) and
superconducting, B$_{1g}$ (blue) and B$_{2g}$ (green) states. We
have assumed that $U_{\bf Q_{AF}}=U_0 \approx 1.66t$. (b) Calculated
dispersion of the resonance peak in CeCoIn$_5$ for the
$d_{x^2-y^2}$-wave symmetry of the superconducting order parameter
along the $(q_x, q_y, \pi/c)$ direction. The abscissa is shown in
units of $\pi/a$. Here, we use $U_{\bf q} \approx U_{{\bf
Q}_{AF}}\left[1-B \cdot \frac{\left({\bf q-Q}_{AF}\right)^2}{{\bf
Q}_{AF}^2}\right]$ with B=0.5.} \label{fig4}
\end{figure}

To address the issue of the resonance peak formation, we show in
Fig.\ref{fig4}(a) the calculated real and imaginary part of the RPA
susceptibility at the antiferromagnetic wave vector in the normal
and the superconducting states.
Among possible superconducting symmetries in CeCoIn$_5$ a resonance
peak forms only for the $B_{1g}$ ( $\Delta_{\bf k} =
\frac{\Delta_0}{2} (\cos k_x a - \cos k_y a)$) symmetry of the
superconducting order parameter. As in the case of CeCu$_2$Si$_2$
the antiferromagnetic wave vector connects states with opposite sign
of the superconducting gap. As is clearly visible from
Fig.\ref{fig3}(a) this results in the formation of a resonance peak
similar to the one in CeCu$_2$Si$_2$. The resonance peak forms near
the particle-hole continuum, {i.e.}, close to $\Omega_c = \min\left(
|\Delta_{\bf k}| + |\Delta_{\bf k+q}| \right)$ which is around
$\Delta_0$. This is because the points connected by ${\bf Q}_{AF}$
are lying relatively far from the part of the Fermi surface where
the gap function has maximum value. Note, if the Cooper-pairing
itself arises due to an exchange of antiferromagnetic spin
fluctuations, the maximum of the superconducting gap occurs at
points of the Fermi surface which are connected by {\bf Q}$_{AF}$.
At the same time, the symmetry of the superconducting gap possesses
still the $d_{x^2-y^2}$-wave symmetry, although with higher
harmonics included. It is remarkable that like in CeCu$_2$Si$_2$ we
find that only a gap function of $d_{x^2-y^2}$-wave ($B_{1g}$) type
results in the formation of the resonance peak at ${\bf Q}_{AF}$.
This unambiguously confirms the bulk symmetry of the superconducting
gap in CeCoIn$_5$. The $d_{xy}$-wave ($B_{2g}$) symmetry discussed
in the literature is clearly ruled out.
%

Finally, in Fig. \ref{fig4}(b) we show the dispersion of the
resonance excitations away from the {\bf Q}$_{AF}$. We observe that
the resonance disperses downwards as a function of frequency. This
behavior is similar to that found in hole-doped high-T$_c$ cuprates.
In particular, the value of the critical frequency $\Omega_c =
|\Delta_{\bf k}| + |\Delta_{\bf k+q}|$ lowers for ${\bf q} < {\bf
Q}_{AF}$ for the in-plane momentum, since the scattering occurs for
states closer to the diagonal part of the Brillouin zone where the
superconducting gap is smaller. As a result, the resonance condition
shifts to lower energies. Our results show that the resonance peak
possesses a universal dispersion away from the wave vector that
connects the maxima of the superconducting gaps and we suggest to
investigate experimentally the dispersion of the resonance.

In conclusion, we analyze the dynamical magnetic susceptibility in
CeCu$_2$Si$_2$ and CeCoIn$_5$ below superconducting transition
temperature. We show that in both cases a resonance feature evolves
at the wave vector of the magnetic instability. Our results show
that the latter two heavy-fermion superconductors and the high-T$_c$
cuprates possess the same symmetry of the superconducting order
parameter suggesting that the same mechanism of the Cooper-pairing
is probably involved. Furthermore, despite of the three-dimensional
electronic structure, the two-dimensional $d_{x^2-y^2}$-wave
superconducting gap in CeCu$_2$Si$_2$ and CeCoIn$_5$ may provide
further hints on the microscopic mechanism of unconventional
superconductivity in heavy-fermion systems as well as in layered
cuprates.

\begin{acknowledgments}
We would like to thank O. Stockert, Ch. Geibel for useful
discussions.  I.E. acknowledges support from Volkswagen Foundation.

\end{acknowledgments}

\end{document}